\documentclass[twocolumn,
superscriptaddress,amsmath,amssymb,
 aps,prl]{revtex4-1}
\usepackage{graphicx}% Include figure files
\usepackage{bm}% bold math

\usepackage{wrapfig}
\usepackage{amssymb, amsmath}
\usepackage{enumitem}
\usepackage{xcolor}
\usepackage{lipsum}
\begin{document}

%\preprint{APS/123-QED}

\title{Shape-morphing dynamics of soft compliant membranes for drag and turbulence modulation}% Force line breaks with \\

\author{Varghese Mathai}
\email{vmathai@umass.edu}
\affiliation{Department of Physics, University of Massachusetts, Amherst, Massachusetts 01003, USA}

\affiliation{Center for Fluid Mechanics, Brown University, Providence, RI 02912, USA}

\author{Asimanshu Das}
\affiliation{Center for Fluid Mechanics, Brown University, Providence, RI 02912, USA}

\author{Dante L. Naylor}
\affiliation{Department of Physics, University of Massachusetts, Amherst, Massachusetts 01003, USA}

\author{Kenneth S. Breuer}
\affiliation{Center for Fluid Mechanics, Brown University, Providence, RI 02912, USA}
\date{\today}

\begin{abstract}
We study the kinematics and dynamics of a highly compliant membrane disk placed head-on in a uniform flow. With increasing flow velocity, the membrane deforms nonlinearly into increasingly parachute-like shapes. These aerodynamically elongated materials exhibit a modified drag law, which is linked to the elastohydrodynamic  interactions. We predict the unsteady structural response of the membranes using a nonlinear, aeroelastic model -- in excellent agreement with experimental measurements of deformations and force fluctuations. With simultaneous membrane interface tracking, force measurements and flow tracing, we reveal that a peculiar skewness in the membrane's oscillations triggers turbulence production in the wake, thereby modulating the drag. The present work provides a demonstration of the complex interplay between soft materials and fluid turbulence, leading to new, emergent system properties. 

%In concurrence with this, we detect that the onset of large amplitude, resonant fluctuations which coincide with turbulence enhancement in the wake. We used particle image velocimetry measurements, combined with an axisymmetric drag model reveals that the origin of the enhanced drag response can be traced back to the turbulence enhancement in the wake.   Through a simultaneous quantification of the unsteady deformations, forces and the flow field, we detect that the onset of large amplitude, resonant fluctuations coincide with turbulence in the wake, which contributes to an increased drag. 
\end{abstract}

\maketitle

%\tableofcontents

%\section*{Introduction}
The interaction of elastic structures with fluids is a problem of central importance to the mechanics of continua and interfaces. When a flexible structure is placed in a flow, its shape change can induce modified interactions between the structure and surrounding flow \cite{alben2015flag,argentina2005fluid,guttag2017active, vogel1984drag, harder2004reconfiguration, alben2004flexibility,vliegenthart2006forced,shelley2011flapping,ganedi2018equilibrium,shelley2005heavy,ristroph2008anomalous,bagheri2012spontaneous,boulogne2022measurement}. This coupling can lead to complex, fluid-structure interactions; common examples range from the fluttering of a flag or a flexible structure in the wind, to the swimming of fish \cite{albenappendage,manela2017hanging,connell2007flapping,hu2008flexible,kim2013flapping,anderson2008physics,manela2017hanging,zhang2000flexible,taguchi2015experimental,schulman2017liquid}. Often, these interactions affect the thrust/drag response of the systems involved \cite{alben2004flexibility,waldman2017camber,alben2008flapping,argentina2005fluid,vogel1984drag,ganedi2018equilibrium,shelley2011flapping,shelley2005heavy,jung2006dynamics,xu2014ultrathin,box2020dynamic,leclercq2018reconfiguration}. A few studies have focused on the steady and unsteady interactions of elastic materials \cite{mavroyiakoumou2020large,tzezana2019thrust}, wherein the materials were typically operated within the linear elastic limit of small strains (e.g., \cite{shelley2005heavy,jin2019flow,tsipropoulos2021interaction}). In contrast, a broader class of highly deformable (nonlinear) materials can be envisioned, with a complex, strain-dependent, elasto-fluidic response. In such situations, the interplay between the nonlinearities of the material and the nonlinearities in the flow could pave way for emergent system properties.

A circular disk placed head-on in a uniform stream represents a classic example of a bluff body flow that has been extensively studied \cite{hoerner1965fluid}. Ganedi et. al. \cite{ganedi2018equilibrium}  recently studied how an oil film suspended by a circular ring deformed in an external flow. When large stretchability, coupled with strain-stiffening behavior (or strain softening), is introduced to this problem, the resulting system can exhibit rich variability in its dynamical behavior. %Ganedi et. al. \cite{ganedi2018equilibrium}  recently studied how a suspended liquid film was deformed by an external flow to form a bubble. The oil film, with its constant surface tension, was observed to bulge due to the competing effects between hydrodynamic pressure and surface tension, and the authors offered a quasi-static explanation for the observed deformations and forces. However, not much attention was paid to the temporal dynamics of the film.  While the liquid film can be considered as the simplest of deformable materials -- with its constant interfacial tension -- more generally, a wider class of soft materials can be envisioned, with a nonlinear, strain-dependent elastic response (Fig.~\ref{fig1:Expt_setup}a,b). 
The unsteady behavior in such situations emerges out of interactions between the material's oscillations and the induced flow field around it.

 \begin{figure}[t]
	\centering{
		\includegraphics[width=0.5\textwidth]{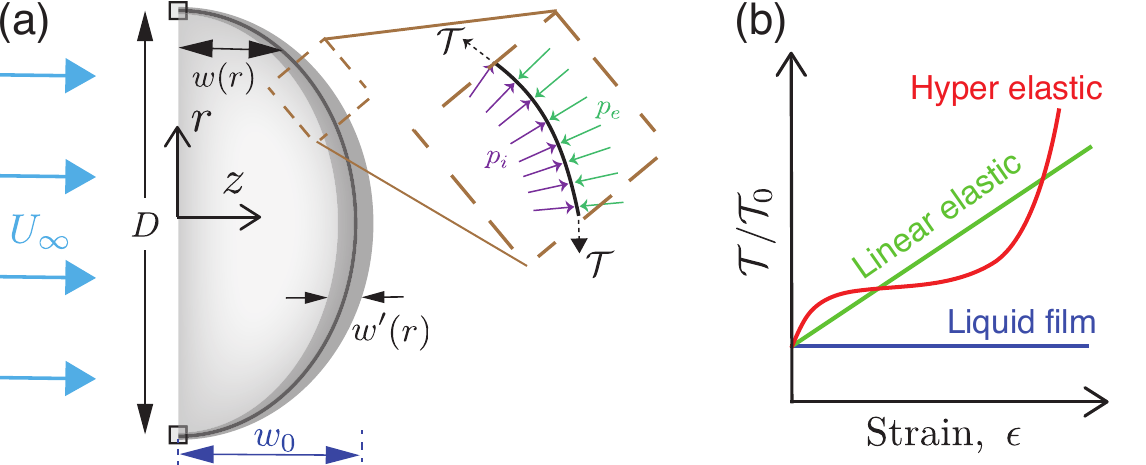}}
	\caption{ (a) Schematic of the side-view of an initially circular membrane disk of diameter, $D$, deforming in response to a uniform incoming flow of velocity, $U_\infty$, {\color{black} where $r$ and $z$ are the radial and axial coordinates, respectively}. The membrane disk is placed inside a low-speed wind tunnel and the membrane bulges to {\color{black}a mean maximum deformation $w_0$, where $w(r)$ is the deformation profile of the membrane. The two squares at the top and bottom denote the cut-section of the rigid circular 
ring that is used to hold the pre-stretched membrane.} The unsteady oscillations induced about the mean bulge are denoted as $w'$. The inset depicts a small portion of the membrane with a force balance {\color{black} between pressures, $p_i$ and $p_e$, and tension $\mathcal{T}$ developed across the membrane} due to stretching (quasi-static approximation). (b) A representative schematic showing the normalized tension ($\mathcal{T}/\mathcal{T}_{0}$) with increasing strain rate ($\epsilon$) for different classes of materials.}
	\label{fig1:Expt_setup}
\end{figure}
 
In this work we present a combined experimental, theoretical, and numerical study of the unsteady fluid-structure interactions of an ultrasoft, compliant membrane placed head-on in a uniform flow. The incoming fluid flow deforms the membrane into parachute-like shapes (Fig.~\ref{fig1:Expt_setup}a). We will quantitatively show how these aeroelastically {morphed membranes} can enter a state of skewed resonance, triggering a modified drag/thrust response when compared to similarly shaped rigid shells. Our analysis, which combines theoretical predictions with membrane interface tracking and time-resolved flow field tracing, reveals that the unsteady motions of the soft, elastic membrane can induce turbulence in the far wake.
 
The membranes were fabricated using an addition-cure type of silicone rubber material with a range of thicknesses, $h = 250 - 3500$ $\mu$m. The shear modulus, $G$, ranged between $4 - 34$ kPa, controlled by adding different amounts of thinner \cite{das2020nonlinear}. A circular sample of the cured membrane was mounted with a desired pre-stretch,  $\lambda_0$, onto a rigid acrylic ring with inner diameter $D = 120$ mm, outer diameter of 128 mm, and thickness 3 mm (see Supplemental Material for details \cite{supp}). 

The membrane disk was fixed to a non-intrusive steel ``claw'', attached to a six-axis load cell and mounted head-on in  the uniform air stream, in a  low-speed wind tunnel, with a test-section of 1.2 m $\times$ 1.2 m cross section, and 3.6 m length. {\color{black} Tests were conducted over a range of flow speeds, $U_{\infty} =  8-25 $ m/s (Reynolds number, $Re \equiv U_\infty D/\nu = 10^5 - 10^6$). The membrane's centerline deflection, $w_0$, was varied in uniform steps from $w_0/D =$ $0.08 - 0.5$ by adjusting the flow speed, and force and torque data were collected at 20 kHz at each of the mean deflection values.} A high-speed camera recorded side-view images at 500 frames/s. In a separate experiment, conducted in a different wind tunnel (test section: $0.61^2$m), velocity fields were measured, at 700 Hz, using Particle Image Velocimetry (PIV) (see Supplemental Material for details \cite{supp}). 

\begin{figure}[!tbp]
	\centering
	\includegraphics[width=0.42\textwidth]{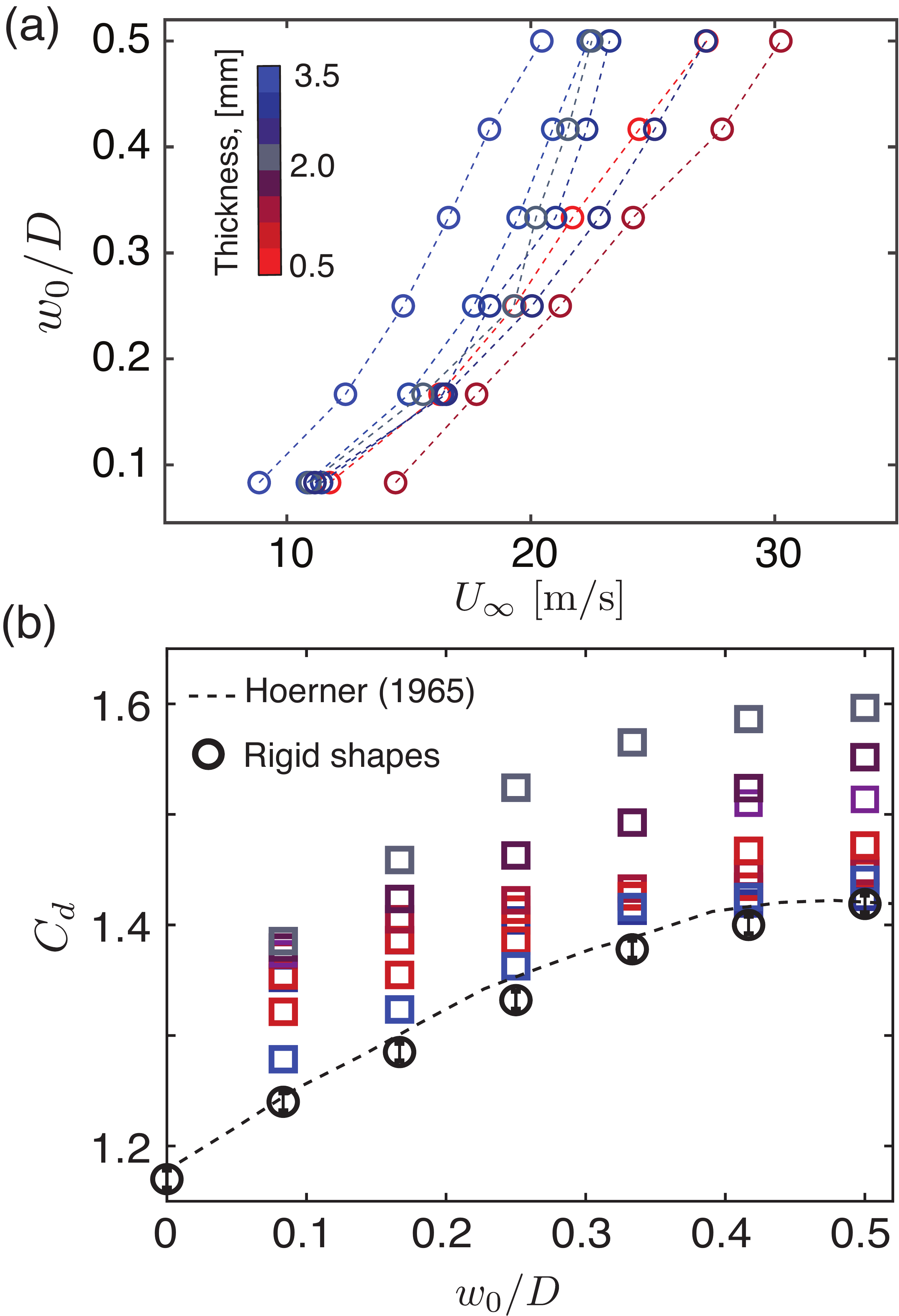}  
		\caption{(a) Normalized mean deformation, $w_0/D$, of various membranes as a function of the flow velocity, $U_\infty$. {\color{black}Each of the dashed lines represents one value of thickness of membrane. The corresponding shear moduli values were $G$ = [1.7, 1.9, 2.1, 3.5. 5.4, 9.5, 19.1] kPa.} (b) Drag coefficient, $C_d$ of various membranes as a function of mean deformation. The color map denotes the membrane thickness. Both the plots demonstrate the wide scatter (non-monotonic) in the deformations and drag measurements.} 
    \label{fig2:Steady_variation}
\end{figure}

As the velocity increases, the membrane starts to balloon from a flat disk shape toward increasingly parachute-like shapes, with a maximum deformation, $w_0$, at the centerline. The resulting steady state deformations (Fig.~\ref{fig2:Steady_variation}a) show dependency on all of the experimental parameters: $G$, $h$, $U_\infty$, and $\lambda_0$, and varies monotonically, but nonlinearly with flow speed. At all deformations the membrane shape is well-approximated by a spherical cap (Supplemental Material \cite{supp}), with curvature,
\begin{equation}
    {\kappa}=\frac{16{w_0}}{D^2+4{w_0}^{2}}.
\end{equation}

The corresponding drag coefficient, $C_d = F_d/(0.5\rho U_\infty^2 A)$, where $F_d$ is the drag force and $A$ is the projected area of the disk, is shown in Fig.~\ref{fig2:Steady_variation}b. The drag coefficient for 3D-printed \emph{rigid} shells (Fig. \ref{fig2:Steady_variation}b; black circles) increases monotonically from a value of 1.17  to a value of $1.42$, in agreement with prior work \cite{hoerner1965fluid}. In contrast, the membranes with the same \emph{mean shape} experience a higher drag. The drag coefficient for the membranes with the same shape varies non-monotonically with the membrane thickness (Fig.~\ref{fig2:Steady_variation}b). Furthermore, when compared to a rigid spherical cap the soft membranes exhibit oscillations about the mean shape (see Supplemental Movie \cite{supp}). 

We can understand the membrane behavior using a simple analytical model, the unsteady deformation of a membrane, ${w}({r},t)$, is given by \cite{smith1995computational}
\begin{equation}
     \rho_m h \dfrac{\partial^2{w}}{\partial t^2} +\mathcal{T} \kappa = \Delta p,
     \label{eq:general-membrane-equation}
\end{equation}
where $\rho_m$ is the membrane's mass density, $\Delta p(r,t)$ is the pressure difference across the membrane, and $\mathcal{T}$ is the membrane tension. Non-dimensionalizing Eq. \ref{eq:general-membrane-equation} using length scale $D$, time scale $D/U_\infty$, and pressure scale  $0.5 \rho U_\infty^2$, where $\rho$ is the fluid density, we obtain
\begin{figure*}[!htp]
	\centering
	\includegraphics[width=1\textwidth]{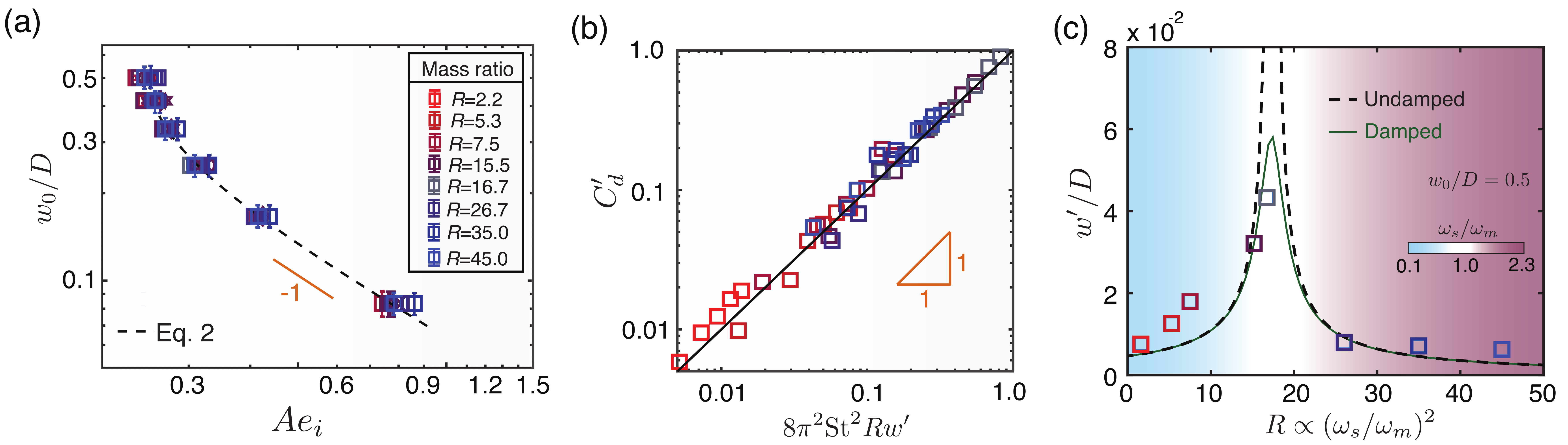}
		\caption{ (a) Comparison of steady state membrane deformation from experiment with the model predictions. Here, the normalized mean deformation is plotted against the effective aeroelastic number, $Ae_i$. In the small deformation limit the model approaches a slope of -1. (b)  Normalised drag fluctuation measurements vs. prediction based on inertial scaling. (c) Amplitude of membrane oscillations, $w'/D$ vs. membrane mass ratio, $R$, for a fixed mean deformation ($w_0/D = 0.5$). The dotted line shows the prediction of resonance from the analytical model (Eq. \ref{eqn:Circular_membrane_force}).  The solid green line incorporates the membrane damping.}
    \label{fig3:model_steady_unsteady}
\end{figure*}

\begin{equation}
\label{eq:non-dimensional-membrane-equation}
      {R} \dfrac{\partial^2 {w}^*}{\partial t^{*2}} + {Ae_i}\ \kappa^* = C_p,
\end{equation}
where $w^*$ and $\kappa^*$ are the dimensionless deformation and three dimensional curvature, respectively, $R = {2\rho_m h}/{\rho D}$ is the mass ratio, $Ae_i ={\mathcal{T}}/({0.5\rho U_{\infty}^2 D})$ is the so-called Aeroelastic number \cite{waldman2017camber}, and $C_p$  is the pressure coefficient (see also Supplemental Material  \cite{supp}).

While Eq. \ref{eq:non-dimensional-membrane-equation} appears to be linear, the membrane's material response and curvature introduce nonlinearity into the second term. The silicone material exhibits a hyper-elastic stretch-strain response and strain-stiffens at large deformations \cite{das2020nonlinear}. Using a two-parameter Gent model for biaxial deformation \cite{gent1996new,das2020nonlinear}, the tension in the membrane can be expressed as $\mathcal{T} = G_m h (1-\lambda^{-6})$, where $G_m={G J_m}/({J_m-I_1+3})$, $G$ is the material shear modulus, $J_m$ is the locking parameter and $I_1$ is the first invariant of the left Cauchy-Green deformation gradient tensor \cite{bower2009applied}. Further, the stretch-ratio for the spherical cap geometry can be written in terms of the curvature and pre-stretch as $\lambda = ({4\lambda_0}/\kappa^*)\sin^{-1}({\kappa^*}/4)$. Combining these, we can express the tension, $\mathcal{T}$, or the aeroelastic number, ${Ae_i}$, in terms of the membrane curvature $\kappa^*$ and other material and fluid properties in Eq.~\ref{eq:non-dimensional-membrane-equation}. 

At steady state, Eq.~\ref{eq:non-dimensional-membrane-equation} yields a relation for $w_0/D$ in terms of $Ae_i$, which needs to be solved implicitly for the entire range of deformation (Fig. \ref{fig3:model_steady_unsteady}a). In the small deformation limit (or large $Ae_i$), the solution can be approximated as
\begin{equation}
    \frac{w_0}{D}  \approx \frac{1}{16 Ae_i}.
    \label{eq:def_small_linear}
\end{equation}
The measured mean deformation, $w_0/D$, is in excellent agreement with the model predictions for all values of $Ae_i$, $\lambda_0$ and $R$. Additionally, uniaxial and biaxial characterizations were conducted in order to model the material stresses in response to prescribed strains (see Supplemental Material for details \cite{supp}).

We turn our attention to the unsteady kinematics of the membrane.  The origin of these fluctuations can be linked to vortex shedding, which is commonly observed to occur in flows over bluff bodies, with a characteristic frequency, $\omega_s$ at a constant Strouhal number, ${St}= \omega_s D/2 \pi U_\infty$ \cite{roshko1961experiments,bearman1984vortex}. The shedding generates an unsteady force, and one can expect the membrane to experience an inertial reaction force, $F_d' = m_m a_m'$, where $m_m$ and $a_m'$ are the membrane mass and characteristic scale of membrane acceleration, respectively. The spectra of the force measurements at all speeds show a nearly constant Strouhal number, $\text{St} =0.12$ (see Fig.~S-1). The acceleration of the oscillating membrane can be expected to scale as $a_m' \propto w' \omega_s^2$, where $w'$ is oscillation amplitude. Therefore, we can express $F_d'=\rho_m h A w' \omega_s^2$, and correspondingly a fluctuating drag coefficient, $C'_d \approx 8 \pi^2 St^2 R w'$. Comparing the experimental measurements of the force fluctuations with this inertial prediction (Fig.~\ref{fig3:model_steady_unsteady}b), we observe an excellent agreement. This demonstrates that the measured drag fluctuations (second moment of $w'$) are predominantly due to the {\it breathing mode} of membrane oscillations (first mode).

Considering the oscillating membrane as a dynamical system, forced at the vortex shedding frequency, we adopt a forced harmonic oscillator model to understand the oscillation amplitude, and to explain the non-monotonic variation of $w'$ with $R$. By invoking axisymmetry, and considering small oscillations about a mean shape, we linearize Eq.~\ref{eq:general-membrane-equation} to obtain
\begin{equation}
    \rho_m h \frac{\partial^2{w'}}{\partial{t}^2} - 2\mathcal{T}\frac{\partial^2{w'}}{\partial{r}^2}= {C_s} F_\text{dyn} \sin\omega_s t.
    \label{eqn:Circular_membrane_force}
\end{equation}
Here the pre-factor $C_s$ is the relative strength of the unsteady vortex shedding forces with respect to the dynamic pressure. A typical bluff body experiences unsteady vortex-induced forces that are about 10\% of $F_{dyn}=0.5\rho U_\infty^2 A$ \cite{mei1991unsteady,halse1997vortex}. Measurements of the force fluctuations for a rigid hemisphere yield $C_s\sim 0.1$ (see Supplemental Material  \cite{supp}). 

The unsteady membrane equation (Eq.~\ref{eqn:Circular_membrane_force}) supports modes that resonate when the natural frequency of the membrane, $\omega_m$, coincides with the frequency of the vortex shedding, $\omega_s$. Approximating that the membrane oscillates similar to a stretched drum (small curvature), the first mode of Eq. \ref{eqn:Circular_membrane_force} has a natural frequency, $\omega_m = 3\pi c/2D$, where $c = \sqrt{2\mathcal{T}/\rho_m h}$ is the wave speed \cite{meirovitch2010fundamentals}. This can be re-written in terms of the Aeroelastic parameter and the mass ratio:
\begin{equation}
    \frac{\omega_m D}{2 \pi U_\infty} =  \frac{3}{2 \sqrt{2}}\sqrt{\frac{Ae_i}{R}}. 
    \label{eq:membrane_frequency}
\end{equation}
Resonance will occur when $\omega_s/\omega_m = 1$, a prediction that is confirmed in our experimental measurement shown in Fig.~\ref{fig3:model_steady_unsteady}c, for $w_0/D = 0.5$. (See also Supplemental Material  \cite{supp}). By measuring the amplitude decay of an oscillating membrane (a ``ring-down'' test - see Supplemental Material  \cite{supp}), we can include an empirical damping term to Eq.~\ref{eqn:Circular_membrane_force}, which provides an upper bound on the amplitude at resonance (solid green curve in Fig.~\ref{fig3:model_steady_unsteady}c). {\color{black} We observe a nearly parameter-independent-resonance point, i.e. a single physical membrane (with $R = 18 \pm 2$) can resonate at a broad range of flow conditions. This has been achieved because the membrane passively adapts its shape and natural frequency in proportion to the change in the flow speed (see Supplemental material \cite{supp} for further details).}

Note the subtle asymmetry observed in the measurements of $w'/D$ about the resonance point (Fig.~\ref{fig3:model_steady_unsteady}c). The origin of this asymmetry lies in the nonlinearity introduced by finite curvature of the membrane -- unaccounted in the simplified drum head model -- which is captured numerically by solving the unsteady membrane structural equation at large oscillation amplitudes (see Supplemental Material \cite{supp} for details).
\begin{figure}[t]
	
\begin{minipage}[b]{0.5\textwidth}
	\includegraphics[width=\textwidth]{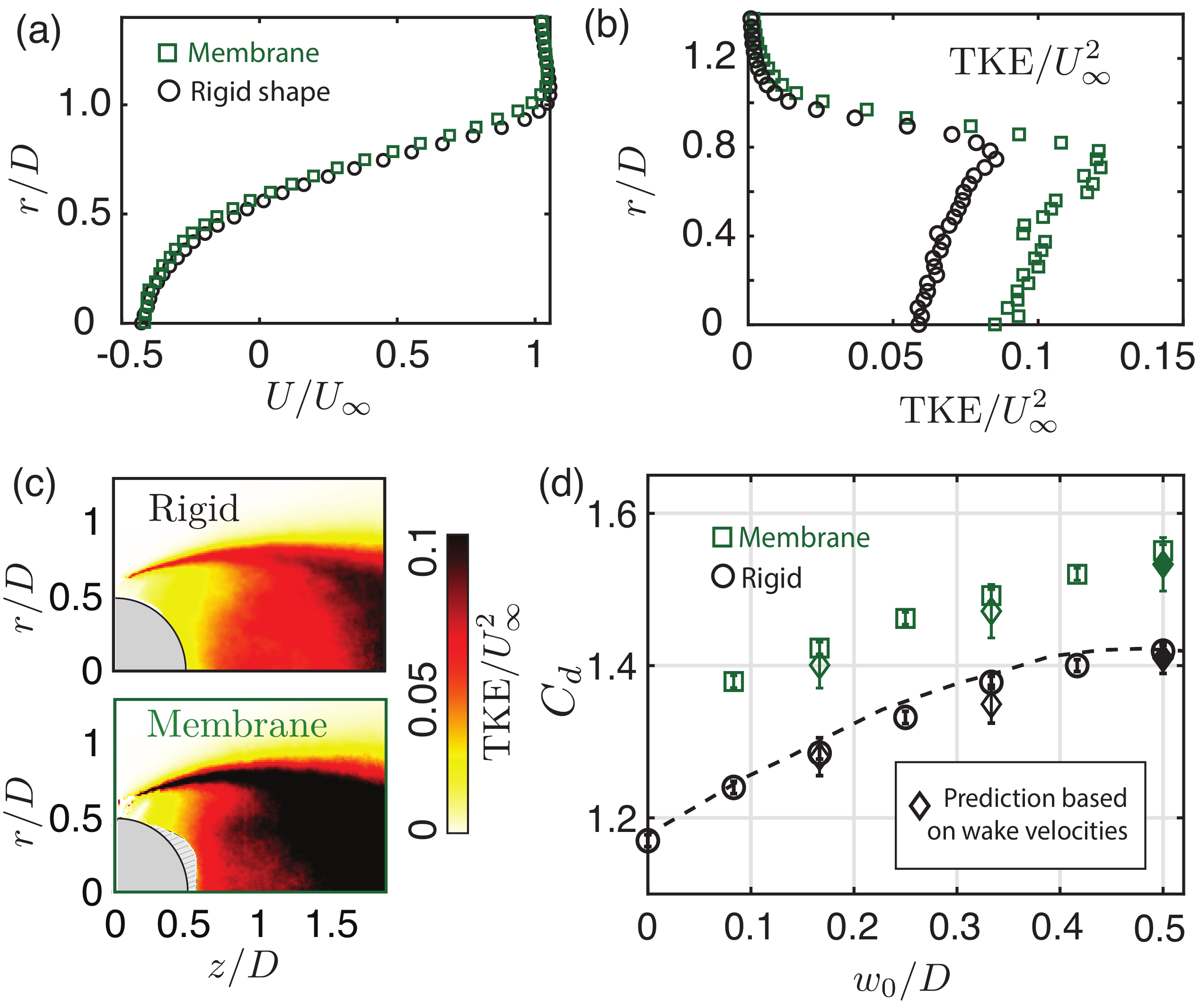}
\end{minipage}
\caption{Comparison of the (a) mean axial velocity profile $\bar{u}_z(r)$ and (b) turbulent kinetic energy (TKE), at $z/D$ = 1.5 downstream of the body with $w_0/D = 0.5$. The TKE field (c) downstream of a rigid shell and deformed membrane disk, with $w_0/D = 0.5$. The hatched region represents an area where velocity vectors were not available due to laser light reflections.  (d) Comparison between the drag coefficient calculated independently from (i) direct force measurements (circles and squares) and (ii) a control volume analysis based on the velocity field (diamond symbols). The black and green symbols are for the rigid shells and the membranes, respectively.}
\label{fig:4_TKE_U}
\end{figure}
Lastly, with the mean deformation and unsteady oscillations explained, we focus on the mechanism responsible for the modified mean drag coefficient, $C_d$, for the membranes. Despite the relatively low oscillation amplitude of the membrane: $w'/D \sim \mathcal{O}(10^{-2})$, the drag coefficient for the membrane is noticeably higher (by up to $20\%$) than that of a similarly shaped rigid shell (Fig.~\ref{fig2:Steady_variation}b). {It is interesting to note that such small amplitudes of oscillations could induce a significant drag modification.}

%The role of membrane vibrations on drag and drag fluctuations is clear from the membrane kinematics and force measurements. 
%Presumably, the drag modulation arises as a result of the coupled interactions of the membrane with the surrounding fluid. 

The drag on the body is reflected in the wake momentum deficit which can be obtained by radial integration of the mean and unsteady wake momentum contributions (see Supplemental Material  for details \cite{supp}):
{ \begin{equation}
   { C_d =\frac{16}{D^2} \int_0^R \Bigg( \ {  \underbrace{\frac{\bar{u}_z}{U_{\infty}}\bigg[1-\frac{\bar{u}_z}{U_{\infty}}\bigg]}_{mean} +  
    \underbrace{{\frac{-\overline{{u'_z}^2}}{U_\infty^2}+\frac{1}{2}\frac{\overline{{u'_r}^2}}{U_\infty^2}}}_{unsteady} \ \Bigg) r dr }}
    \label{eq:axi_drag_coeff}
\end{equation}}
We perform two-dimensional particle image velocimetry (PIV) of the wake behind the membrane and a similarly-shaped rigid shell, measuring the axial,  $u_z$,  and radial,  $u_r$,  velocities. Comparing the velocity fields from these two cases, we find that the mean wake velocity profiles, $\bar{u}_z$,  are nearly identical (Fig.~\ref{fig:4_TKE_U}a), and hence the contribution to the drag from the steady term in Eq.~\ref{eq:axi_drag_coeff} is comparable for the two cases. However, the turbulent kinetic energy, TKE~$\approx 3/4(u_z'^2+u_r'^2)$, in the wake behind the membrane is significantly greater than for the rigid shell (Fig.~\ref{fig:4_TKE_U}b,c), and 
when one includes the unsteady velocity terms in the calculation of $C_d$, we find excellent agreement between the force measurements and the PIV estimations for both the membranes and the rigid shells (Fig.~\ref{fig:4_TKE_U}d). Downstream in the wake, the small-scale fluctuations are expected to tend toward local isotropy, and the periodic signature of vortex shedding has nearly disappeared \cite{davoodianidalik2022fluctuation}. Remarkably, the increase in the wake TKE exceeds the energy density of the oscillating membrane by an order of magnitude, i.e. ${u_z'^2/(w'\omega_s)^2} \sim \mathcal{O}(10)$ and it is this energy that accounts for the increase in the mean drag coefficient. The weak correlation between TKE production and $w'$ can be rationalized by noting that it is the subtle skewness of oscillations that drives turbulence production. We performed a set of interface-resolved numerical simulations of a membrane oscillating within a fluid flow field. The membrane's rate of stretching as compared to its relaxation rate, i.e. the skewness of motion, dictates the degree of drag modulation (drag increase vs. drag reduction; see Supplemental Material \cite{supp}). A detailed exploration of this is part of an ongoing investigation.

In summary, we have conducted a systematic study of the aeroelastic response of an ultrasoft membrane disk in a uniform flow. We observe that the material deforms nonlinearly into parachute-like shapes. The time-averaged shape of the membrane can be accurately modeled using a hyperelastic Gent constitutive model \cite{gent1996new} that depends on a single dimensionless parameter - the Aeroelastic number, $Ae_i$.  The unsteady membrane vibrations are driven by vortex shedding, and the fluctuations are accurately modeled using a simple spring-mass system that depends on the Aeroelastic number and the membrane mass parameter, $R$.  {\color{black}Through shape-morphing, the membrane adapts its natural frequency with the flow speed, resulting in a single physical membrane exhibiting (or avoiding) resonance over a broad range of flow conditions.} We anticipate that triggering the nonlinear elastic response of materials within fluid flows may open up a number of opportunities for drag control using soft, stretchy materials.
\vspace{0.1 cm}
~\\
We thank Anupam Pandey and Detlef Lohse for fruitful discussions. K.B acknowledges funding from the U.S. Army/Soldier Systems Center, Natick, MA. and support from NSF Grant \#2035002. D.L.N. acknowledges funding from the Kenneth and Joanne Langley Research Fund and the Simenas Fellowship.

V. M. and A. D contributed equally to this work and are joint first authors. Experiments by A. D. and V. M. Numerical simulations and theoretical work by V. M., D. L. N., A. D., and K. B. Data analysis and writing of the manuscript by V. M., K. B. and A. D. Project conception by K. B and V. M.

%\bibliography{reference_prl}

\providecommand{\noopsort}[1]{}\providecommand{\singleletter}[1]{#1}%

\end{document}